# Coopetition of software firms in Open source software ecosystems

Anh Nguyen Duc[1], Daniela S. Cruzes[2], Geir K. Hanssen[2], Terje Snarby[3]
and Pekka Abrahamsson[1]

[1]Norwegian University of Science and Technology
{anhn, pekkaa}@ntnu.no
[2]SINTEF Digital
{geir.k.hanssen, daniela.s.cruzes}@sintef.no
[3]Genus AS
terjesnarby@gmail.com

**Abstract.** Software firms participate in an ecosystem as a part of their innovation strategy to extend value creation beyond the firm's boundary. Participation in an open and independent environment also implies the competition among firms with similar business models and targeted markets. Hence, firms need to consider potential opportunities and challenges upfront. This study explores how software firms interact with others in OSS ecosystems from a coopetition perspective. We performed a quantitative and qualitative analysis of three OSS projects. Finding shows that software firms emphasize the co-creation of common value and partly react to the potential competitiveness on OSS ecosystems. Six themes about coopetition were identified, including spanning gatekeepers, securing communication, open-core sourcing and filtering shared code. Our work contributes to software engineering research with a rich description of coopetition in OSS ecosystems. Moreover, we also come up with several implications for software firms in pursing a harmony participation in OSS ecosystems.

## 1 Introduction

Increasingly, software products are no longer developed solely in-house, but in a software ecosystem (SECO), where developers collaborate with "*distributed collaborators*" beyond their firm boundary [1, 12]. This differs from traditional outsourcing techniques in that the initiating actor does not necessarily own the software produced by contributing actors and does not hire the contributing actors. All actors, however, coexist in an interdependent way. Game developers in App Stores, for instance, might share a similar game engine, but independently produce different applications to mobile users. By integrating with SECOs, firms can benefit from developing projects of a size that exceeds their own capabilities, exploring opportunities to enter new markets [14], performing a inside-out process [2], and employment of a recruitment strategy [15].

Before the full potential advantages of SECOs are leveraged, commercial firms need to consider several concerns. At the organizational level, the firm's benefit and the ecosystem's goal are not always the same [3]. Participation of commercial firms in SECOs with their diverse motivations and business strategies might introduce dynamics, and sometimes conflicts in navigating the project evolution [14]. The body of knowledge in open source software (OSS) projects provide sufficient amount of knowledge on firms' motivation, collaboration patterns, and business models when participating in such an open collaborative firm network [4, 5, 6, 7]. However, one often-neglected aspect is the consideration of both competition and collaboration among firms, as two sides of the same coin.

Coopetition, as a concept, relates to the coexistence of competition and collaboration, and conceptualizes the interaction among firms with a partial congruence of interests [8, 9]. In a coopetitive environment, firms cooperate with each other to reach a higher value creation as compared to the value created without interaction and struggle to achieve competitive advantage. A good example of coopetition in a restaurant business is when a large number of restaurants are concentrated in a relatively small area ("*the restaurant district*" or "*the restaurant quarter*"). Coopetition takes place when companies' being in the same market work together in the exploration of knowledge and research of new products. Since coopetition applies to inter-firm relationships, open source SECO offers an ideal context for understanding coopetition among firms that develop and utilize a common software codebase [11].

Our research objective is to explore the state-of-practice on coopetition among commercial firms in open SECOs. To our knowledge, there exist only a few studies that examine the coopetition phenomenon between commercial firms in SECOs [10, 11, 14], making it an interesting research topic. A research question (RQ) was derived from this research objective:

> *RQ: How do commercial firms maintain both collaboration and competition in an open source software ecosystem?*

The study is organized as follows: Section 2 presents a background about coopetition and firm participation in open source SECOs. Section 3 describes our research methodology, Section 4 presents our findings, and Section 5 discusses the findings. Finally, Section 6 concludes the paper.

## 2   Background and Related Work

### 2.1. Coopetition among software firms

Coopetition, as a business management concept, conceptualizes the interaction among firms in relation to their strategic development [8, 9]. Dagnino et al. is among the first authors that proposed a definition of coopetion as a new way to capture inter-firm dynamic interdependence, which includes both cooperative and competitive perspectives [26]. The authors proposed two forms of coopetition, dyadic coopetition (concerns among two-firm relationships) and network coopetion (involving more than two firms, i.e. value chain) [26]. Our case represents a simple network coopetition, which is described by coopetition among multiple firms at the same level of a value chain.

In general, coopetion is a complex yet important phenomenon that is worth further research [28]. Coopetition is also considered as an important element for linking between R&D and production within the firms. By selecting an OSS project with high innovation that provides technical advantages to a software firm, we investigate a suitable case for building an understanding of coopetition in the software industry.

There exist few empirical studies about coopetion among software firms [10, 11, 14]. Valenca et al. explored the concepts of competition and collaboration in requirement engineering processes [14]. The authors investigated two firms that participated in a collaborative network evolving towards a SECO. The firms faced challenges in requirement negotiation and lack of sufficient coordination with the common project. The authors conclude that even though competition is inevitable among companies, establishing long-term partnership are crucial drivers for innovation and performance. Our study, however, investigates the coopetition at the implementation stage instead of the requirement stage.

The more relevant work to our study is from Bengtsson et al. [9], Teixeira et al. [11, 29] and Linaker et al. [10], by exploring how rival firms collaborate in an OSS project using data mining and social network analysis techniques. Teixeira et al. observed a different result compared with traditional management literature, stating that competition for the same business model does not necessary affect collaboration within the SECO. Bengtsson et al. argued that developers within a firm need to be divided to take charge of either collaboration or competition [9]. Linaker et al. investigated the changing stakeholder influence and collaboration patterns in the Hadoop project [10]. The authors highlighted that independent of business model, all firms work together towards the common goal of advancing the shared platform [10]. Our study complements to these findings, but also bring new understanding about coopetition via a comprehensive research approach.

### 2.2. Firm participation in software ecosystem

Multiple definitions of a SECO exist [15], while we refer to the one by Jansen et al. [12], as "*a set of actors functioning as a unit and interacting with a shared market for software and services, together with relationships among them. These relationships are frequently underpinned by a common technological platform or market and operates through the exchange of information, resources and artifacts.*" Manikas performed a literature review on recent SECO research [15], describing social characteristics of SECO, i.e. geographical distribution and management of engineering practices [11, 17, 18, 19]. There are also empirical studies about actors-to-actors dependencies and relationship, such as software supply networks [20, 21], collaboration patterns among SECO actors [22, 23].

The influence of firm participation in OSS communities has been studied from different angles, leading to different observations. Mehra et al. showed that the heterogeneity, which existed between firm-paid developers and voluntary developers shaped the evolution of OSS community and product [24]. Dahlander et al. studied the network of relationships within the GNOME project, discovering that the presence of hired developers often generates an initial diffidence among unpaid programmers [25]. Lamastra et al. found that firm's involvement improved the ranking of OSS projects, but, on the other hand, lowers software quality, probably because of corporate constraints put on the OSS developing practices [13]. These studies provide a basis for understanding firm participation in OSS, as well as possible methodological approaches to explore the topic.

## 3  Research Approach

### 3.1. Study design

We conducted this work by using a multiple-case study design [27]. Exploratory case studies are suitable to explain the presumed causal links in real-life interventions. There are abundant OSS projects available, many of them are abandoned or individual efforts. We are interested in OSS projects which are large enough and impactful. A brainstorm session was conducted among the paper's authors and an external collaborator to decide the case selection criteria:

- Commercial participation: the selected case should have multiple commercial firms participating in the development. In addition, there must be an adequate way to identify them.
- Successful and on-going: the OSS project must be successful and on-going. This implies that the project attracts developers and the development of the software is progressing.

- Active projects with many activities: There must be a high level of communication and code commits in the project.

As a result, we came up with a list of possible projects that satisfy all criteria. Two projects, that we found most relevant to our research were selected, namely Wireshark and Samba. Wireshark[1] is an OSS toolkit developed by a community of networking experts around the world under the GNU General Public License. The project is officially operated under the Wireshark name since May 2006. Out of the 802 developers listed in Wireshark contributor list, 342 were classified as firm-paid developers (43%). The remaining 460 developers (57%) were classified as volunteering developers. The firm contributions come from 228 firms. Samba[2] is an OSS suite that provides file, print and authentication services to all clients using the SMB/CIFS protocol. Samba is licensed under the GNU General Public License, and the Samba project is a member of the Software Freedom Conservancy. In Samba, 316 developers were evaluated, where 182 (57%) of them were classified as firm-paid developers. The contributions come from 45 firms. Communication and collaboration between developers in the Wireshark and Samba community mainly occur in two places; the developer mailing list and the bug tracking system.

Later, a third OSS project was selected following the same criteria, in order to (1) update the project sample, which might be aging and (2) provide complementary qualitative data. Bootstrap[3] is a frontend Javascript-based framework for developing responsive, mobile first projects on the web. The project was released as an OSS project since 2011. At the time the research is conducted, Bootstrap is the most-starred project on GitHub, with over 90 thousands stars and more than 38 thousands forks. Source code and issue management is done via Github. The communication in Bootstrap was done via many channels, i.e. StackOverflow, Slack, and Github tracker. Besides studying available document and project infrastructure, we were able to interview three developers in the Bootstrap project.

### 3.2. Data collection and analysis

The main data collection process occurred between Sep 2012 and May 2013. During this phase, both quantitative and qualitative data was collected. Complementary data was collected between April 2015 and August 2015. The main source of quantitative data is from mailing lists, code and issue repositories, as they are common data sources when studying OSS [4, 10, 19, 22]. The main qualitative data comes from semi-structured interviews with firm-paid developers in Wireshark and Samba.

---

[1] https://www.wireshark.org
[2] https://www.samba.org
[3] http://getbootstrap.com

**Table 1: The most crowded firms participating to Wireshark and Samba**

| Wireshark | | | Samba | | |
|---|---|---|---|---|---|
| Firm | # of devs. | % of devs. | Firm | # of devs | % of devs |
| Cisco | 16 | 2% | IBM | 17 | 5,4% |
| Ericsson | 11 | 1,4% | RedHat | 14 | 4,4% |
| Siemens | 8 | 1% | SerNet | 8 | 2,5% |
| Netapp | 6 | 0,7% | SUSE | 8 | 2,5% |
| Citrix | 5 | 0,6% | EMC | 4 | 1,3% |
| Lucent | 5 | 0,6% | SGI | 4 | 1,3% |
| MXTelecom | 5 | 0,6% | Exanet | 3 | 0,9% |
| Nokia | 5 | 0,6% | HP | 3 | 0,9% |
| Axis | 4 | 0,5% | Cisco | 3 | 0,9% |
| Harman | 4 | 0,5% | Canonical | 2 | 0,6% |

We decided to extract data from all available project public infrastructures, such as project wiki pages, developer mailing lists (referred to as mailing lists), bug tracking systems and code repositories. We collected developer profiles from public sources of information, such as project wiki and confluence pages. Basic information, like developer email addresses and the time stamp when changes to a specific file had been made can be extracted from JIRA and GIT. The communication data was collected from two main sources, which are bug tracking systems and mailing lists. We used a name and an email address to identify whether a participant is from a firm. The approach has been successfully used to do similar classifications [4, 24]. The top ten firms participating in the OSS projects with regard to number of developers is presented in Table 1. The percentage represents the portion of developers for the referring firm in the total number of project contributors. In Wireshark, only 8 % of the firms have 3 or more developers participating in the community. Whereas, 78 % of the firms have only one developer participating.

Regarding to qualitative data, interviews were selected from a convenient sample consisting of the firm-paid developers from Wireshark, Samba and Bootstrap. As we did not know much about the population, we aimed for a non-probabilistic sampling technique using a conjunction of purposive and snowball sampling. In Wireshark, we used an existing connection to one of the core contributors as a starting point, and asked for suggestion of developers that could be interesting to interview next. The core contributor pointed out relevant developers for the research topic, and assisted in contacting them by posting our interview invitation on the core contributor mailing list. In Samba, we selected relevant developers in the OSS project based on the quantitative data and sent interview invitations to these by email. In Bootstrap, we had a developer actively contributing to the project in our personal network. From him, we got two more interviews with firm-paid participants in Bootstrap.

**Table 2: Summary of interview profiles**

| Alias | Domain | Firm type | Firm size | SECOs |
|---|---|---|---|---|
| D1 | Telecommunication | Corp. | 10 000+ | Wireshark |
| D2 | Wireless networking services | SME | 11 - 50 | Wireshark |
| D3 | Messaging system | SME | 11 – 50 | Wireshark |
| D4 | Telecommunication | Corp. | 10 000+ | Wireshark |
| D5 | IT security services | | 51 - 200 | Samba |
| D6 | Server and OS development | Corp. | 10 000+ | Samba |
| D7 | Telecommunication | Corp. | 10 000+ | Samba |
| D8 | Social media | Startup | 1 - 10 | Bootstrap |
| D9 | Hosting and file sharing | SME | 51-200 | Bootstrap |
| D10 | Social media | Startup | 1 - 10 | Bootstrap |

The interview guide consisted of four to five main topics, with both closed and open questions. The closed questions were mainly used in the introduction phase of the interview to solicit background information about the respondent, firm and OSS project context. In addition, closed questions were used to confirm or attribute statements given by other developers. The open questions were used to collect information about: (1) work process/bridge engineer role, (2) firm awareness/organizational boundary and (3) position in the community/contributions. The interviews were conducted in English, except for one. The duration of the interviews ranged from 45 minutes to 72 minutes. All the live interviews were recorded to facilitate subsequent analysis and minimize potential data loss due to note-taking. These recordings were thereafter transcribed verbatim. Transcribing audio records resulted in 55 pages of rich text.

The analysis of the qualitative data was undertaken following guidelines and recommended steps for thematic synthesis in SE [16]. This thematic analysis approach allows the main themes in the text to be systematically summarized and is also familiar by the first two authors of the paper. A basic outline of the process is illustrated in Figure 1. Segments of text about firms' interaction, i.e. activities, attitudes about communication, collaboration and competition were identified and labeled. After two rounds of reviews of the data, we ended up with 84 codes.

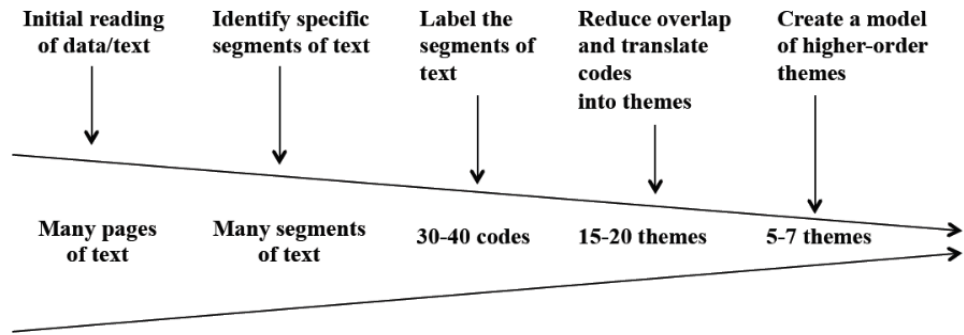

**Figure 1: Thematic analysis [16]**

The following step of the thematic analysis was to translate the codes and the corresponding text segments into themes. A theme in this context is essentially a code in itself, however, a theme is an increased distanciation from the text, and thus an increased level of abstraction. The codes were evaluated and combined to form an overreaching theme, which describes how software firms interact with each other in OSS projects.

## 4 Results

We found six main themes related to coopetition among firms, which are: Organizational boundary spanning via gatekeepers (Section 4.1), Securing communication among actors on firm competitive advantages (Section 4.2), Open-core sourcing policy (Section 4.3), Business driven filtering of code sharing (Section 4.4), Value of social position in OSS community (Section 4.5), and Friendly competitiveness (Section 4.6).

### 4.1. Organizational boundary spanning via gatekeepers

The perceptions of a gatekeeper, who **navigates code and information flow** between his/ her firm and external actors, were acknowledged by all the interviewees (as shown in Figure 2). D1 stated that when his coworkers found issues with the third party components, they informed D1, but not project managers. D7 expressed a similar perception: "*Yes, I act as a bridge between [Firm Name] and Samba and forward bugs/errors to the community*." The gatekeeper is the hub of information and issues that can be reached by different developers across the organizations, as stated by D4: "*Yes, everybody definitely knows that I am the Wireshark guy. All the developers, testers and customer support people know that they can come to me if* *they have Wireshark issues...*". The gatekeeper is often an active actor in contributing to the ecosystem, as mentioned by D2: "*Many of our core developers are working for smaller companies, and have a responsibility for the internal protocols that their company needs. (...) I think most developers work individually, and have the role of providing Wireshark functionality to the other developers in the firm.*" In firms with multiple developers active in upstream development, i.e. committing to OSS projects, there is often a recognized gate keeper role among them. D5 mentioned: "*In general when it comes to contributing patches upstream each developer in [Company Name] is independent and can directly approach the upstream project… The [Company Name] Samba package maintainer usually has a task of being the **gatekeeper** for those bugs that have been reported against [Company Name] products by the customers or the support teams...*" In this case, while code is contributed independently by individuals in the firm, the bugs is managed by a gatekeeper who submits bug reports on behalf of the firm into the OSS project's bug tracking system.

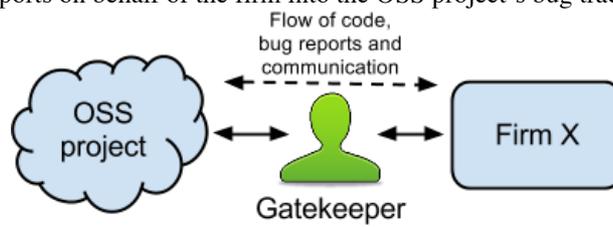

**Figure 2: The role of gatekeeper in a commercial firm**

### 4.2. Securing communication among actors on firm competitive advantages

Among various communication channels in the OSS projects, firms secure communication related to firms' competitive advantages. Communication channels are mainly e-mail and instant messaging, and in some cases Skype and telephone. D3 said: "*I have done it [contacted developers directly] different times in the past. Not just as a general 'I am stuck, can you help', but because it would be an area I knew the other guy was working on.*" D6 mentioned: "*Usually I tend to do R&D tasks myself. I often seek for reviews of my work. When I need assistance, I will go directly to a developer in the community.*" D8 considered private communication as a way to establish **high-quality contact point** and **potential collaboration for further projects**. D9 mentioned: "*We try to address as much as we can of the issues that come to us… Normally if we get a private message about an issue, we will give it higher priority …*". D5 mentioned that when **discussing legal or security sensitive issues** he used a private communication channel. The nature of such issues invokes the use of private channels as posting it in the public channels may result in security breaches or similarly bad situations. Although none of other developers said anything about the use of direct channels for such issues, we believe that it is a common procedure in most OSS projects.

### 4.3. Open-core sourcing policy

Despite of risks and issues with competitors, commercial firms are quite open in sharing and collaborating in their source code. In an open-core approach, firms participating in an upstream approach **contributing all the code** they develop to the OSS project's public sources, and collaborate exclusively within the OSS project to develop the software. D5 described the upstream development approach by his firm: "*In general, our philosophy is to develop upstream first and then back-port changes that have been approved by the upstream community into our products. We stay very involved in the communities and try to keep the differences between our packaged software and upstream software to the minimum necessary.*" One of the expected benefits was **to avoid maintenance and merging issues** when combining public parts of private parts of source codes. D10 illustrated for this idea: "*… if you are to make a change in the core, and you want to keep it private, you will have to fork the project and maintain it yourself. (...) I believe, in the general case, that you gain more from contributing to the development, that retaining your code from the community*".

### 4.4. Business driven filtering of code sharing

Firms contribute code that is (1) related to the core of the OSS projects and (2) code that is regarded as open and/or standardized, and collaborate within the community to develop the code they contribute. Such firms typically have private repositories where they have code related to the OSS which is proprietary and thus retained from the public sources. **Not all the code** that is written in the firm is contributed back to the OSS project. D4 mentioned: "*The majority of the stuff I have written for Wireshark has been pushed up… But you sort of draw a line in the stuff that is obscure enough to not push. The only people who should be looking at our proprietary protocol should be us…*". Some of the code is regarded as proprietary and is retained in the firm's private code repository, due to **technical specific**, or **legal and authorization issues.** D2 mentioned: "*Mainly protocol dissectors for protocols used in our equipment, if the protocol is based on open protocol descriptions from 3GPP, ITU or IETF (RFC) it is considered OK to make an individual contribution to OSS...*"

### 4.5. Value of social position in OSS community

For a firm, the social position in the ecosystem is perceivably useful and important. It is apparent that a central position in the community is closely related to being a core developer in most cases. Two concrete benefits mentioned by the interviewees are: (1) **easier for code inclusion** and thereby avoid the need of having a private code repository, and (2) **receiving more help** from other community members. D1 elaborated the value of his position within the community: "*Researcher: do you think that it is an important position for firms to have in OSS communities? D1: Yes, because when we are doing changes, we can incorporate them into Wireshark pretty quickly. We don't have to maintain our own code base and synchronize it. We just commit code to the source and have it there.*"

D4 highlights the importance of social position in OSS community: "*I think it [having a position] helps a lot. I think there is a difference if, lets say, D2 asks for help, then I'll help him if I can. But if Joe from I have never really heard of, is asking for help then my level of effort is usually lower. And part of that is because I know D2 personally, and part of that is because I know that he does a tremendous amount of work.*"

Firms seem not to utilize their social positions to dominate the OSS development. D6 mentioned: "*Before working on Samba I used to think that big companies may have big influence in OSS projects simply by "buying" core developers. Now, that I know most of the people working on Samba, I know that this is not feasible.*" Hence, having a position, or "*buying*" one, is neither the way firms relate to nor the tactic firms influence the OSS development.

### 4.6. Friendly competitiveness

Firms working within the same business domain are often competitors in the market, and thus it is interesting to see how influential the firm awareness is when firms come together in community based OSS projects to develop software collectively. Surprisingly, the firm-paid developers say that they **perceive** other developers **as partners and/or friends rather than competitors**. D5 pointed out that he had met many of the developers at the developer conferences, and considered many of them as friends. D1 explained that he did not make any distinction between a firm-paid developer and a volunteering developer, and said: "*I think of them as developers, and not about which firms they represent.*" D7 say that he perceives others as partners. D6 mentioned: "*I've always thought of others as partners. Even more - I think about them as colleagues.*" D4, D8 and D9 shared similar thoughts, and dismissed the perception of other firm-paid developers as competitors: "*I guess as things have evolved we do actually compete in some respects with some of these people at this point. But that hasn't really occurred to me much… I have noticed more people who tend to be customers of ours, rather than true competitors. We might be competitors within some areas, but I have never really thought about it I guess*", stated by D9.

The issues of competition from a firm from the other side of the world might not be relevant for a startup and a SME who are **pushing efforts on having their product released**. Without a clear vision on how their market or technical advantages are influenced by sharing and using OSS source code, the concern of competition is not much relevant. D8 also mentioned: "*…you think about other firms as your* *competitors, but I don't think that really comes in to my interactions really. They have their own users somewhere around the world…. I have sometimes seen contributions from their developers, but I think that is good…*"

The firm awareness in the community is perceived as valuable. However, developers remark that it is not the knowledge of what other firms work for that is valuable, rather it is **the knowledge of what business domain** they are working within. D2 replied when was asked about other firm awareness: "*Yes, but I don't know that much about the firms of the other developers. They typically say that they work for Firm X, and that's it. What firm they are working for is not that important to me.*" D3 emphasized the potential value of having the firm awareness: "*… I know that D2 may have some role as a contact for Firm X… I know that D2 may be someone who is good at getting log files for specific things. In the past when I was working with voice over IP, I thought sometimes he was able to give me some log files from within his company, but I didn't really think of him as the company representative. I think of him as a company person who may be able to get logs for me, like he does.*"

Additionally, the interviewees were asked if they considered that their contributions could be used by others firms to gain or recapture competitive advantage. The majority dismissed this perception, for example: "*As Firm X does not directly control Wireshark, I guess we have to be a bit careful when we are in contact with other developers. (…) I believe, in the general case, that you gain more from contributing to the development, that retaining your code from the community*", stated by D2. A final remark by D5 about the competitiveness: "*Although there may be some competition between companies, as engineers we seek collaboration for mutual benefit. We already know any advancement will be used by everybody, that's not a problem, we get back as much as we give out.*"

## 5   Discussion

Table 3 summarizes the identified themes that describe how firms interact with each other in three popular OSS projects. For each theme, we classified whether they belong to a collaborative relationship or a competitive relationship. While some of the concepts are not surprising compared to what is known in OSS research, i.e. social position in OSS community [10, 11, 19], open-core sourcing [2], they are interesting contributions in exploring how software firms manage both collaboration and competition in OSS ecosystems. We also found novel concepts about coopetition, such as securing communication and friendly competitiveness. Interestingly, some phenomenon that we initially thought as competitive activities, turned out to be collaborative, such as gate keepers and friendly competitiveness.

**Table 3: Summary of key findings**

| Themes | Description | Category |
|---|---|---|
| Organizational boundary spanning via gatekeepers | One/ few persons who navigates code commits, Q&A | Collaborative |
| Securing communication among actors on firm competitive advantages | Limited sensitive information to certain partners | Competitive |
| Open-core sourcing policy | Publish all of their code, complete in sync with upstream development | Collaborative |
| Business driven filtering of code sharing | Filtering technical specific, legal, strategic modules | Competitive |
| Value of social position in OSS community | Appreciate the better position in OSS community | Competitive |
| Friendly competitiveness | Attitude of cooperating rather than competing | Collaborative |

Dagnino et al. highlight that coopetition does not simply emerge from joining competition and collaboration, but rather it implies that collaboration and competition merge together to form a new kind of strategic interdependence between firms [26]. Alternatively, our cases show that firms focus on activities that create a common value with an awareness of not sharing their technical and legal sensitive information. Our study reveals the competition mode partly appears at software code level, which is represented by the filtering of code sharing and the open-core sourcing policy. Even when firms are aware of their competitors, the attitude of collaboration is still overwhelming. Valenca et al. raise a question whether firms are collaborators or competitors in SECO context? At the requirement engineering level, the authors found several significant challenges among firms within the same collaborative network [14]. OSS projects and firms might have divergent interests but firms can manage to discover areas of convergent interest and be able to adapt their organizing practices to collaborate [3]. In our case, this is clearly shown at the source code level. The finding also matches with observations by Linåker [10].

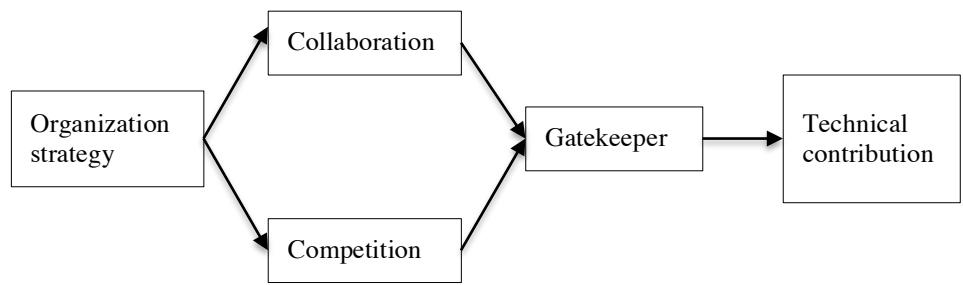

**Figure 3: A model of firm coopetition in open source SECO**

Bengtsson et al. argued that individuals within a firm can only act in accordance with one of the two logics of interaction at a time, i.e., either to compete or to collaborate [9]. Our observation on the gatekeeper role gives a possible alternative theory on how firms manage such coopetition scenario. By influencing the gatekeeper, who manages code flows and information flows between the firm and the SECO, the firm can implement competing or collaborating strategies. The firm strategy can be flexible, for example fully open core sourcing at one time, and filtering of shared code at another time. The implementation of such strategies is done via the firm gatekeeper, who does actual technical contribution to the SECO. Therefore, in contrast with Bengtsson's findings, we argued that it is possible to implement a firm-level dynamic interaction via individuals in software projects, as shown in Figure 3.

We propose a framework of coopetition in open source SECOs, as described in Figure 3. Derived from the firm's strategy when participating in an open source SECO, the firm involves in both collaborative and competitive relationships with other firms participating in the SECO. Balancing and managing both logics of interaction were done by a gatekeeper role, which can be one or a group of key developers that are active in the SECO. The gatekeepers implement coopetition by carrying out different mechanisms (i.e. described in Section 4.1 to Section 4.4), which eventually realizes as technical contributions to the SECO.

## 6 Conclusions

Coopetition is an important foundation for economics and management research [26], but often overlooked or oversimplified in other domains. In SECOs, where inter-firm interactions are crucial for both firms and SECO development, coopetition is a relevant and also a new way of looking at SECOs. Our contributions are two folds (1) we portrayed the situations where both competition and collaboration occurs in OSS projects and (2) we proposed a framework to explain how a gatekeeper could help to manage such coopetition relationship.

For coopetition research, we offer an alternative explanation of how coopetition is performed by software firms in an open source SECO. We observed that software firms emphasized the co-creation of common value and partly react to the potential competitiveness on OSS ecosystems. For Software Engineering research, the work illustrates the adoption of a management theory in understanding and exploring both technical and business aspects of SECOs. Through the lens of coopetition, novel aspects of inter-firm interaction in OSS projects were highlighted. The research contributes to the current body of knowledge on SECO by adding the competition perspective.

For software firms who participate in an open source ecosystem, our findings offer a descriptive insight about different coopetition strategies observed in a community-driven OSS project. Firms can refer to different ways of co-creating via collaboration and awareness of competition when they participate in such an ecosystem. For proprietary SECO steering members, the harmony interaction observing in our open source SECOs can help direct implication on how to design and to influence the SECO policy to support a healthy coopetition.

For future work, the next step would be to refine and to validate the coopetition model (described in Figure 3) with a larger set of cases. Our research here only uses three community-driven OSS projects, which limits the generalization of findings to other types of SECOs, such as proprietary platforms, firm-driven OSS projects, etc. Future work is needed to explore the concept of coopetition in such contexts. Besides, a longitudinal observation on how coopetition evolve among firms can provide knowledge that goes beyond cross-sectional observations. Furthermore, we also plan to triangulate the observations from manager and developer's viewpoint. Last but not least, further investigation about employing the role of gatekeepers for coopetition is needed to validate our observation.

<>*This is the author's version of the work. It is self-arhived at Arxiv. The definite version was published in: Nguyen Duc A., Cruzes D.S., Hanssen G.K., Snarby T., Abrahamsson P. (2017) Coopetition of Software Firms in Open Source Software Ecosystems. In: Ojala A., Holmström Olsson H., Werder K. (eds) Software Business. ICSOB 2017. Lecture Notes in Business Information Processing, vol 304. Springer, Cham, https://doi.org/10.1007/978-3-319-69191-6_10*

28. T. Annika, Causes of conflict in intercompetitior cooperation, *Journal of Business and Industrial Marketing*, vol 24(7), pp. 508-516, 2009
29. J. Teixeira, S. Q. Mian, and U. Hytti, Cooperation among competitors in the open-source arena: The case of OpenStack. ICIS, 2016